\documentclass[aps,amsmath,amssymb,prb,superscriptaddress,twocolumn,showpacs,floatfix]{revtex4}
\usepackage{graphicx}
\usepackage[english]{babel}

\def\etcl{$\kappa$-(BE\-DT\--TTF)$_2$\-Cu\-[N\-(CN)$_{2}$]Cl}

\def\agetcn{$\kappa$-(BE\-DT\--TTF)$_2$\-Ag$_2$(CN)$_{3}$}
\def\etcn{$\kappa$-(BE\-DT\--TTF)$_2$\-Cu$_2$(CN)$_{3}$}

\def\kcucn{$\kappa$-CuCN}
\def\kagcn{$\kappa$-AgCN}
\def\cm{cm$^{-1}$}

\begin{document}
\title{Anions effects on the electronic structure and electrodynamic properties\\
 of the Mott insulator $\kappa$-(BE\-DT\--TTF)$_2$\-Ag$_2$(CN)$_{3}$}
\author{M.\ Pinteri\'{c}}
\affiliation{Institut za fiziku, P.O.Box 304, HR-10001 Zagreb, Croatia}
\affiliation{Faculty of Civil Engineering, Smetanova 17, SI-2000 Maribor, Slovenia}
\author{P. Lazi\'{c}}
\affiliation{Rudjer Bo\v{s}kovi\'{c} Institute, Bijeni\v{c}ka cesta 54, HR-10000 Zagreb, Croatia}
\author{A.\ Pustogow}
\affiliation{1.\ Physikalisches Institut, Universit\"{a}t Stuttgart, %Pfaffenwaldring 57,
D-70550 Stuttgart, Germany}
\author{T.\ Ivek}
\affiliation{Institut za fiziku, P.O.Box 304, HR-10001 Zagreb, Croatia}
\author{M.\ Kuve\v{z}di\'{c}}
\affiliation{Institut za fiziku, P.O.Box 304, HR-10001 Zagreb, Croatia}
\affiliation{Department of Physics, Faculty of Science, University of Zagreb, P.O.Box331, HR-10001 Zagreb, Croatia}
\author{O.\ Milat}
\affiliation{Institut za fiziku, P.O.Box 304, HR-10001 Zagreb, Croatia}
\author{B.\ Gumhalter}
\affiliation{Institut za fiziku, P.O.Box 304, HR-10001 Zagreb, Croatia}
\author{M.\ Basleti\'{c}}
\affiliation{Department of Physics, Faculty of Science, University of Zagreb, P.O.Box331, HR-10001 Zagreb, Croatia}
\author{M.\ \v{C}ulo}
\affiliation{Institut za fiziku, P.O.Box 304, HR-10001 Zagreb, Croatia}
\author{B.\ Korin-Hamzi\'{c}}
\affiliation{Institut za fiziku, P.O.Box 304, HR-10001 Zagreb, Croatia}
\author{A.\ L\"{o}hle}
\affiliation{1.\ Physikalisches Institut, Universit\"{a}t Stuttgart, %Pfaffenwaldring 57,
D-70550 Stuttgart, Germany}
\author{R.\ H\"{u}bner}
\affiliation{1.\ Physikalisches Institut, Universit\"{a}t Stuttgart, D-70550 Stuttgart, Germany}
\affiliation{Inst. Functional Matter and Quantum Techn., Universit\"{a}t Stuttgart, D-70550 Stuttgart, Germany}
\author{M.\ Sanz Alonso}
\affiliation{1.\ Physikalisches Institut, Universit\"{a}t Stuttgart, %Pfaffenwaldring 57,
D-70550 Stuttgart, Germany}
\author{T.\ Hiramatsu}
\affiliation{Faculty of Agriculture, Meijo University, Nagoya 468-8502, Japan}
\author{Y.\ Yoshida}
\affiliation{Faculty of Agriculture, Meijo University, Nagoya 468-8502, Japan}
\author{G.\ Saito }
\affiliation{Faculty of Agriculture, Meijo University, Nagoya 468-8502, Japan}
\affiliation{Toyota Physical and Chemical Research Institute, Nagakute 480-1192, Japan}
\author{M.\ Dressel}
\affiliation{1.\ Physikalisches Institut, Universit\"{a}t Stuttgart, %Pfaffenwaldring 57,
D-70550 Stuttgart, Germany}
\author{S.\ Tomi\'{c}}
\affiliation{Institut za fiziku, P.O.Box 304, HR-10001 Zagreb, Croatia}
\email{stomic@ifs.hr}
\homepage{http://sceinlom.ifs.hr/}

\date{\today}

\begin{abstract}
The Mott insulator $\kappa$-(BEDT-TTF)$_2$\-Ag$_2$(CN)$_3$ forms a
highly-frustrated triangular lattice of $S=1/2$ dimers 
with a possible quantum-spin-liquid state. Our experimental and
numerical studies reveal the emergence of a slight charge imbalance between
crystallographically inequivalent sites, relaxor dielectric response
and hopping dc transport. 
In a broader perspective we conclude that
the universal properties of strongly-correlated charge-transfer salts
with spin liquid state are an anion-supported valence band
and cyanide-induced quasi-degenerate electronic configurations in the relaxed state.
The generic low-energy excitations are caused by charged domain
walls rather than by fluctuating electric dipoles. They give rise to glassy
dynamics characteristic of dimerized Mott insulators, including the
sibling compound $\kappa$-(BEDT-TTF)$_2$\-Cu$_2$(CN)$_3$.
\end{abstract}

% Infrared and Raman spectra

\maketitle

%\section{Introduction}
\label{sec:Intro}
Electronic ferroelectricity and multiferroicity attracts great attention of condensed matter physicists due to their fundamental and technological importance.\cite{Ikeda2005,Kimura2008,Ishihara2010}
They are identified in systems with strong electronic correlations
such as transition-metal oxides and low-dimensional charge-transfer molecular solids.
In the latter category, electric polarization arises from valence instability and charge ordering.
In both cases, breaking the inversion-symmetry results in the concurrence of non-equivalent charge-sites and bonds.\cite{vandenBrink2008}
There is no doubt that electron correlations are fundamental for stabilizing the ferroelectric ground state, nevertheless, experimental evidence indicates
that the delicate interplay of Coulomb forces and structural changes within the coupled molecular-anion system have to be taken into account.
Along these lines a solid understanding of electronic ferroelectricity was achieved for the families of quasi-one-dimensional organic charge-transfer salts: (TMTTF)$_2X$ and TTF-$X$, but also some layered (BEDT-TTF)$_2X$ systems.\cite{Monceau2012,deSouza2013,STMD2015}

However, no consensus has been reached yet on the origin of the ferroelectric signatures
detected in the strongly dimerized $\kappa$-(BE\-DT\--TTF)$_2X$ salts.\cite{Abdel-Jawad2010,Lunkenheimer2012,Tomic2013,ItohPRL2013,Pinteric}
In these compounds, the BEDT-TTF dimers are arranged in a triangular lattice with a relatively high geometrical frustration.
In some of them, indications of charge-ordering phenomena have been reported, but in-depth studies are missing \cite{Swietlik2007,Drichko2014}. On the other hand, the Mott dimer insulators \etcl{} and \etcn{}, called \kcucn{}, have been thoroughly studied because they are discussed as prototypes of a molecular multiferroic and quantum spin liquid (QSL) systems.\cite{Lunkenheimer2012,ShimizuPRL2003}
It turns out to be extremely challenging to reconcile the
idea of quantum electric dipoles on molecular dimers interacting via dipolar-spin coupling\cite{Hotta,Ishihara,Sumit,Gomi}
with the experimentally evidenced absence of any considerable charge imbalance.
So far no global structural changes and no charge disproportionation between molecular dimer sites
larger than $2\delta_{\rho} \approx \pm 0.01e$ that could break the symmetry have been found.\cite{Sedlmeier2012,Kanoda2012}
In the case of the QSL \kcucn{} extensive numerical and experimental investigations\cite{DresselRC2016} have recently revealed that the inversion-symmetry
is broken on a local scale due to linkage isomerism in the anion layer;
the relaxed structure exhibits the low P1 symmetry.
Absorption bands in the THz frequency range -- previously attributed to collective excitation of intradimer electric dipoles\cite{ItohPRL2013} -- are traced back to coupled anion-dimer vibrations.
These findings support our interpretation that the relaxor-like dielectric behavior in the kHz-MHz range resembles the dynamics due to charge defects in the random-domain structure of \kcucn.\cite{Pinteric}

\begin{figure}%FIG1
\centering
%\vspace*{-7mm}
\includegraphics[clip=true,width=\columnwidth]{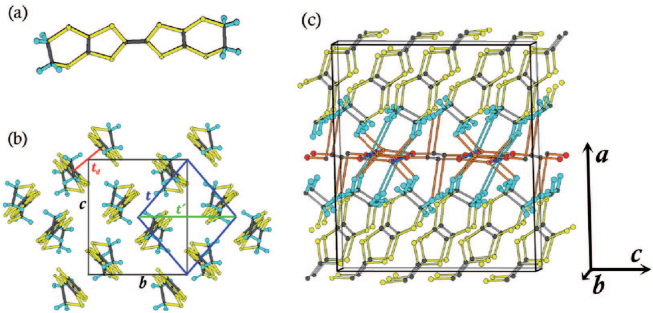}
%\vspace*{-10mm}
\caption{(Color online)
(a) Sketch of the bis(ethyl\-ene\-di\-thio)\-te\-tra\-thia\-ful\-va\-lene molecule, called
BEDT-TTF.
(b) View of BEDT-TTF dimers in the $bc$ plane projected along the direction tilted from the $a$-axis by $42^\circ$; an almost isotropic triangular lattice is denoted by full thick lines; the interdimer transfer integrals are labeled by $t$ and $t^{\prime}$, while the intradimer transfer integral is labeled by $t_d$; the unit cell is denoted as a rectangle. (c) Side view of extended unit cell of $\kappa$-(BE\-DT\--TTF)$_2$\-Ag$_2$(CN)$_{3}$. Possible hydrogen bonds between the CH$_2$ endgroups of the BEDT-TTF molecules and CN$^-$ groups of the anion network are indicated by full lines.}
\label{fig:structurecomposite}
\end{figure}

In order to elucidate the ferroelectric nature of QSL and advance the understanding of its dielectric response, the newly synthetized Mott dimer insulator \agetcn,\cite{Hiramatsu2016} called \kagcn{}, has drawn our attention as recent NMR experiments have indicated a QSL ground state.\cite{Shimizu2016} In this compound (Fig.\ \ref{fig:structurecomposite}) the anions also affect the overall electronic structure, but to a lower  degree compared to \kcucn.\cite{DresselRC2016}
We have carried out dielectric, dc resistivity and vibrational spectroscopy measurements and supplemented them by density functional theory calculations.
In both systems the relaxed structure consists of quasi-degenerate electronic states related to different cyanide configurations; however, in \kagcn{} the higher-symmetry space groups P$2_1$ and P$_c$ are preserved and we resolve a slight charge inequality
between two non-equivalent sites. Still, the universal nature of the low-energy charge excitations in the QSL of $\kappa$-(BE\-DT\--TTF)$_2X$ systems is described more appropriately by mobile charged domain walls rather than by fluctuating electric dipoles.
The dielectric response is due to cooperative process between electronic correlations and local scale symmetry breaking in the coupled molecular-anion system that are present already at 300\,K.

%\section{Experimental and results}

\begin{figure}%FIG2
\includegraphics[clip,width=0.9\columnwidth]{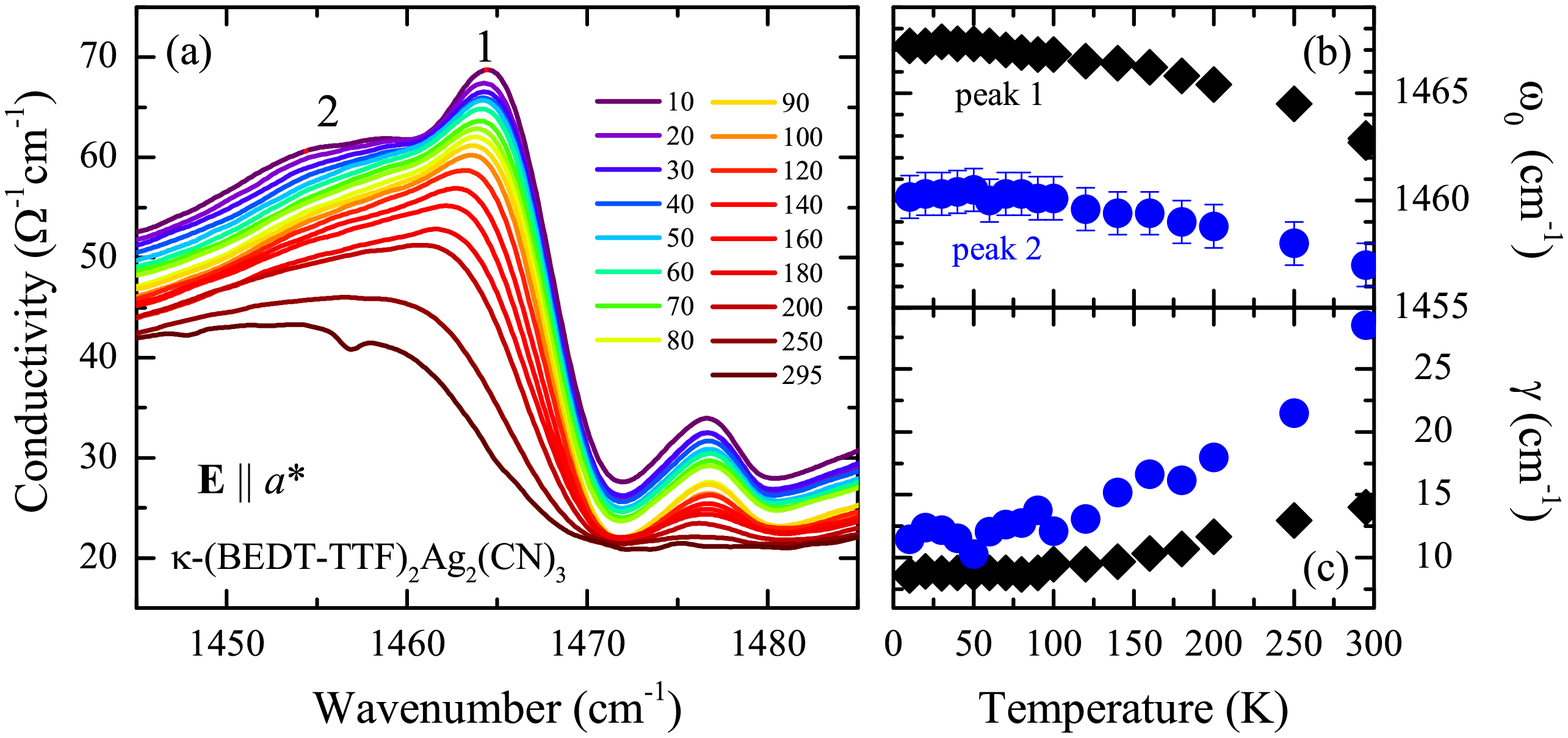}
\caption{(Color online)(a) Temperature evolution of the intramolecular vibration $\nu_{27}$ in $\kappa$-(BE\-DT\--TTF)$_2$\-Ag$_2$(CN)$_{3}$. The two Fano functions required to fit the spectra indicate two crystallographically distinct sites.
No effect on cooling rate (0.1 and 1~K/min) was observed. Temperature dependence of the (b) resonance frequency and (c) damping.}
\label{fig:nu27}
\end{figure}

Let us start by looking at the molecular level and in particular the charge per BEDT-TTF.
Vibrational spectroscopy was performed on high-quality \kagcn\ single crystals\cite{Hiramatsu2016}
by an infrared microscope attached to the Fourier-transform interferometer.  In Fig.~\ref{fig:nu27} the most charge-sensitive intramolecular vibrational mode $\nu_{27}$, which involves the anti-symmetric ring C=C stretching of the BEDT-TTF molecule,\cite{Dressel04,Yamamoto05,Girlando11}
is plotted for various temperatures. No splitting of the mode occurs on lowering the temperature down to 10\,K, providing clear evidence that no sizable static charge redistribution takes place in \kagcn; the observations are in full accord with previous reports on \kcucn.\cite{Sedlmeier2012} In the latter compound, however, only one Fano mode was sufficient to describe the $\nu_{27}$; for \kagcn\ two Fano functions are required, as becomes obvious by the two peaks in Fig.\ref{fig:nu27}(a). They are separated in frequency by 6~\cm\ independent on temperature within the uncertainty [(Fig.\ \ref{fig:nu27}(b)];
we interpreted it as two unequal sites in the unit cell.

\begin{figure}[b]%FIG3
\includegraphics[clip,width=\columnwidth]{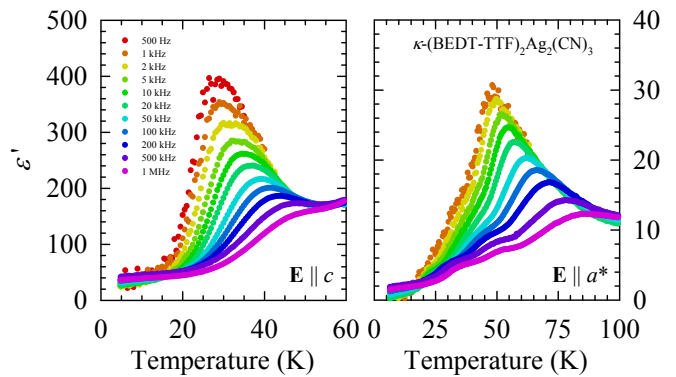}
\caption{(Color online) Real part of the dielectric function $\varepsilon'$ of $\kappa$-(BE\-DT\--TTF)$_2$\-Ag$_2$(CN)$_{3}$ as a function of temperature. The frequency dependent measurements performed with the ac electric field applied  $\mathbf{E}\parallel c$ (left panel) and $\mathbf{E}\parallel a^\ast$ (right panel) both demonstrate the relaxor-ferroelectric behavior.}
\label{fig:Tsweeps}
\end{figure}

Long-wavelength collective excitation are probed by dielectric spectroscopy combined with
dc-transport measurements carried out along the $c$ and $b$ axes within the molecular planes and along the $a^\ast$-axis perpendicular to them.
In the frequency range 40\,Hz--10\,MHz the dielectric function was obtained from the complex conductance measured in the temperature sweeps between 300 and 4.2\,K using several
precision impedance analyzers. Fig.~\ref{fig:Tsweeps} displays the temperature-dependent real  part of the dielectric function at selected frequencies along the in-plane ($c$-axis) and along the out-of-plane ($a^\ast$-axis) directions: with lower frequencies
the magnitude of the peak increases and its position shifts to lower temperature.
A pronounced peak structure with a strong frequency dependence is observed that obeys Curie's law,
indicating a gradual freezing of the dielectric response.
The behavior demonstrates a ferroelectric-like response in the presence of disorder,
similar to the observations in \kcucn.\cite{Abdel-Jawad2010,Pinteric}

\begin{figure}%FIG4
\includegraphics[clip,width=0.5\columnwidth]{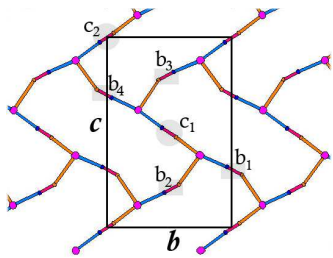}
\caption{(Color online) Relaxed structure of the Ag$_2$(CN)$_3^-$ network in the $bc$ plane projected along the $a$-axis as obtained by DFT calculations. In this configuration, the bridging cyanides at the center (c$_{1}$) and at the corner (c$_{2}$) of the unit cell are aligned antiparallel and directed along the [021] direction. In the each layer, two segregated chains can be identified in which cyanides b$_{1}$ and b$_{2}$, and b$_{3}$ and b$_{4}$ are arranged in the opposite directions, respectively. Silver is colored in red; carbon and nitrogen of CN groups are colored in blue and orange. The unit cell is marked as a rectangle.}
\label{fig:anion}
\end{figure}

In order to understand the observed dielectric and vibrational features, we have carried out {\it ab-initio} calculations in the framework of density functional theory (DFT) as implemented in the VASP code using the projector augmented-wave method; the phonon modes were obtained with PHONOPY code via the DFT-derived force constants.\cite{Kresse93,Blochl94,Kresse99,Bignozzi94,Togo08}
Our numerical calculations are based on the x-ray data taken at 150\,K, solved in P2$_1/c$ and reduced to P1.\cite{SM}
The whole \kagcn{} system relaxes into a structure of minimum energy
when the bridging cyanides in anion network Ag$_2$(CN)$_3^-$ at the center
and at the corners of the unit cell are aligned antiparallel,
whereas the cyanides in the segregated chains are arranged in the opposite directions (Fig.\ \ref{fig:anion}). The relaxed structure is characterized by the space group P2$_1$; meaning that while the two (BEDT-TTF)$_2^+$ dimers within the unit cell remain symmetrically equivalent, the two molecules within each dimer are inequivalent.

\begin{figure}%[h]%FIG5
\includegraphics[clip,width=0.8\columnwidth]{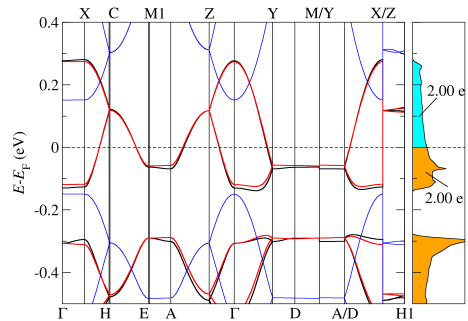}
\caption{(Color online) Band structure of $\kappa$-(BE\-DT\--TTF)$_2$\-Ag$_2$(CN)$_{3}$ (black) plotted along the high-symmetry directions. 
If only the molecular subsystem is considered, the generic cation bands (blue) are obtained. Shifting them by 424\,meV (red) results in an almost complete coincidence with the band structure of the whole system, thus revealing the overall cation-derived band character around the Fermi level. The occupation of the corresponding density of states indicates a integer charge transfer for half-filled band, as expected from the chemical composition.
}
\label{fig:band_structure}
\end{figure}

The corresponding band structure of \kagcn\ is presented in Fig.~\ref{fig:band_structure}. Its overall features
resembles the \kcucn-analogue;\cite{Nakamura09,KandpalPRL2009,DresselRC2016} in particular the cation-derived character around the Fermi level is supported by the rigid band shift of the cation subsystem alone. Here, the shift and bandwidth are smaller by 76\,meV and 20\,meV, respectively,
and the full charge transfer of 2 electrons takes place,
in contrast to only 1.93 for \kcucn.\cite{DresselRC2016}
Thus, the signature of silver $d$-orbitals in the band structure and the density of states is less pronounced. These particularities can be attributed to the different level of hybridization for silver and copper $d$-orbitals: the former is located further below the Fermi level by additional 2\,eV.
Of equal importance is the higher symmetry preserved in \kagcn{} due to a weaker perturbation by the anion layer.

In P2$_1$ symmetry, the energy difference between two possible electronic configurations
(antiparallel and reversed antiparallel alignment of bridging cyanides) is only 25~meV.
Interestingly, also the two electronic states with P$_c$ symmetry \cite{note0} related to two remaining bridging cyanide orientations are only 5--16 meV higher than the relaxed state with minimum energy.\cite{note1}
These results imply that the electronic ground state is quasi-degenerate,
leading to a random domain structure of \kagcn\ in real space.
Note also that flipping one cyanide in the chain  costs 320~meV and increases to 1220~meV for all four CN groups; this process is possible but rather unlikely.

Our findings demonstrate that relaxor ferroelectricity is due to the quasi-degenerate nature of the electronic ground state of \kagcn.
The degeneracy is triggered by breaking the inversion-symmetry on a local scale
because the isomorphism in the cyanides results in a random domain structure.
Importantly, the higher symmetries P2$_1$ and P$_c$ prevail in \kagcn, but not in \kcucn{}
where all bridging CN configurations induce low symmetry P1.
Our x-ray data verify these numerical results: they contain reflections forbidden for the high-sym\-metry group P2$_1/c$ (Fig.~S1), although it remains valid on the global scale.\cite{Shimizu2016}
Another strong evidence is provided by vibrational spectroscopy: the $\nu_{27}$ feature contains two Fano modes as opposed to only one in \kcucn.
We explain this observation by a higher level of disorder related to P1 symmetry.

The reason for this distinctly different behavior lies in the arrangement of the BEDT-TTF dimers with respect to the anion layers. As pointed out by Saito,\cite{Saito} in the case of \kcucn\ the dimers sit exactly in the opening of the hexagon defined by the anions, while for \kagcn\ the center of the dimer is located on the top of the rim of the anion opening (Fig.~S2).
The long contacts between the bridging CN and the terminal ethylene of the BEDT-TTF lead
to a weaker perturbation of the molecular layers in \kagcn{} in contrast to \kcucn{}
where all contacts are shorter than the sum of van der Waals radii.

\begin{figure}[b]%FIG6
\includegraphics[clip,width=0.9\columnwidth]{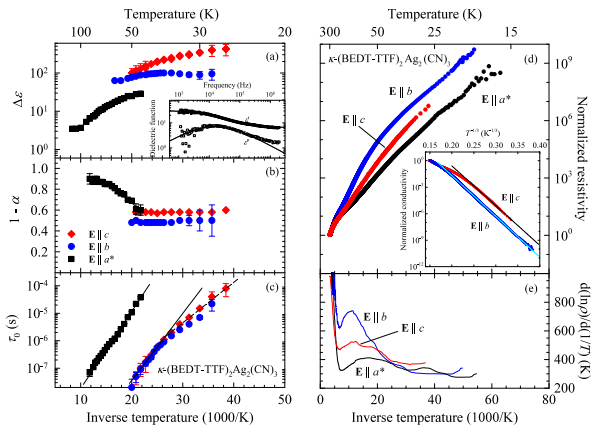}
\caption{(Color online) (a) Dielectric strength $\Delta\varepsilon$, (b)~distribution of relaxation times ($1-\alpha$), (c) mean relaxation time $\tau_0$ of $\kappa$-(BE\-DT\--TTF)$_2$\-Ag$_2$(CN)$_{3}$ as a function of inverse temperature.  The inset of (a) shows a double logarithmic plot of the frequency dependence of the real $\varepsilon'$ and imaginary $\varepsilon''$ parts of the dielectric function for $\mathbf{E}\parallel a^\ast$. The full lines are fits to a generalized Debye function $\varepsilon(\omega)-\varepsilon_{\mathrm{HF}} = {\Delta\varepsilon}/[{1+({\rm i}  \omega \tau_0)^{1-\alpha}}]$; $\varepsilon_{\mathrm{HF}}$ is the high-frequency dielectric constant. 
In (c) the full and dashed lines indicate the Arrhenius slowing down of the primary and secondary relaxation process, respectively. (d)~DC resistivity normalized to room temperature value and (e)~logarithmic resistivity derivative versus inverse temperature. The inset displays the normalized dc conductivity as a function of $T^{-1/3}$: the behavior evidences the variable-range hopping in two dimensions, whereas above about 140\,K, the behavior crosses over into the nearest-neighbor hopping described by a simple activation. }
\label{fig:dielectric_dc}
\end{figure}

The lower disorder level in \kagcn{} compared to \kcucn{} is not a question of crystal quality but rather represents an intrinsic property. It results in fewer domains of larger size in the Ag-compound.
Domains are related to different cyanide configurations separated by charged domain walls.
These interfaces extend to the molecular layers via hydrogen bonds of the BEDT-TTF molecules.
By this mechanism we can explain the dielectric response in \kagcn{} where fingerprints of a glass ordering process are revealed as shown in Fig.~\ref{fig:dielectric_dc}(a,b,c);
similar observations are reported for \kcucn.\cite{Abdel-Jawad2010,Pinteric}

Concomitantly the lower level of disorder corresponds to reduced in-plane dielectric strength $\Delta\varepsilon$ in \kagcn\ compared to the Cu-analogue.
The relaxation time distribution is very broad.
Upon cooling the system freezes:
the mean relaxation time follows an Arrhenius behavior.
For $\mathbf{E}\parallel a^\ast$ we can extrapolate to a value of 100~s
at the glass transition $T_g \approx 22\,K$.
Within the planes, this primary process is detected in a rather narrow frequency and temperature window only while another, secondary process, which sets in at about 36\,K becomes clearly visible.
The contribution of charged domain walls, given by $\Delta\varepsilon$,
is enhanced as the temperature is lowered
because screening becomes weak due to the reduced number of charge carriers.
Consequently, it takes much longer for the system to respond.
At low temperatures this primary relaxation process freezes out
and is replaced by the secondary process involving rearrangements over the lower energy barriers.

We can also identify the influence of heterogeneities in the dc transport
that takes place by hopping within the molecular planes. As demonstrated in
Fig.~\ref{fig:dielectric_dc}(d) variable-range hopping in two dimensions
is clearly revealed in a wide low-temperature range: $\sigma(T) \propto \exp\left[-(T_0/T)^{1/(d+1)}\right]$ with $d = 2$.
At elevated temperatures a crossover to nearest-neighbor hopping is observed with an activated behavior
$\sigma(T) \propto \exp(-\Delta/k_B T)$.\cite{Mott1971} We estimate $k_B T_0 = 95$~eV  and $\Delta=0.08$~eV.
Looking at the derivative of $\rho(T)$ plotted in Fig.\ \ref{fig:dielectric_dc}(e), two features can be identified.
The first concerns the right wing of the peak centered at 300\,K, which is attributed to ordering of the the ethylene group into eclipsed conformation.\cite{Shimizu2016}
Secondly, broad maxima are identified at about 85 to 65\,K
for all directions within and perpendicular to the planes.
This feature may indicate the formation of a low-temperature phase; the width of the peak suggests
that it develops only on short length scales.
It is interesting to note that in the same temperature range
changes in the vibrational features are observed.
As illustrated in Fig.~\ref{fig:nu27}(b,c),
the resonance frequencies and the damping parameters of two Fano modes level off below 80~K,
exactly where the dc resistivity exhibits the broad maximum in its temperature derivative.

Let us return to the broad $\nu_{27}({\rm b}_{1u})$ vibration \cite{SM} (Fig.~\ref{fig:nu27}).
The two maxima separated by 6~\cm\ correspond to an average charge imbalance of
$2\delta_{\rho} \approx \pm 0.05e$.\cite{SM,note3}
While this value is too small to evidence real charge order,
it indicates a different charge per molecule within a dimer for the P2$_1$ and a variation of average charge between dimers for P$_c$ configurations.
As we saw above, the BEDT-TTF molecules are affected by the randomness of the bridging cyanides and
this is why their nature, static or fluctuating, cannot be identified uniquely.
Given the extremely small charge imbalance and fast rate of oscillations,
such fluctuating dipoles cannot explain the low-frequency dielectric response.
Instead we identify the mobile charged domain walls
as the low-energy excitations responsible for the observed relaxor dielectric response.

%\section{Conclusion}
In conclusion, we have combined DFT calculations with dielectric, transport, optical, and structural measurements of the highly frustrated organic dimer Mott insulator $\kappa$-(BEDT-TTF)$_2$\-Ag$_2$(CN)$_3$
and compared the findings with \etcn.
We demonstrate that the ground state consists of electronic states quasi-degenerate in energy,
which reflect a random domain structure.
Charged mobile domain walls rather than fluctuating electric dipoles are the generic low-energy excitations responsible for the relaxor-ferroelectric response in these quantum-spin-liquid systems.
While the overall properties of these two compounds  are rather similar, we are able to
identify some differences in symmetry and microscopic domain structure due to distinct interaction between anions and BEDT-TTF molecules. More comprehensive x-ray diffuse scattering and diffraction measurements at low temperatures are indispensable for a full understanding of these strongly correlated organic systems.

%\begin{acknowledgments}
Technical assistance in x-ray diffraction measurements of D.\ Matkovi\'{c}-\v{C}alogovi\'{c} and I. Halasz are gratefully acknowledged. The work has been supported by the Croatian Science Foundation project IP-2013-11-1011. We appreciate financial support by the Deutsche Forschungsgemeinschaft (DFG) and Deutscher Akademischer Austauschdienst (DAAD), as well as by the Japan Society for the Promotion of Science (JSPS) KAKENHI Grant Number JP23225005.
%\end{acknowledgments}


\begin{thebibliography}{99}



\bibitem{Ikeda2005}
N. Ikeda, H. Ohsumi, K. Ohwada, K. Ishii, T. Inami, K. Kakurai, Y. Murakami, K. Yoshii, S. Mori, Y. Horibe and H. Kito, Nature {\bf 436}, 1136 (2005).

\bibitem{Kimura2008}
T. Kimura, Y. Seiko, H. Nakamura, T. Siegrist and A. P. Ramirez, Nat. Mater. {\bf 7}, 291 (2008).

\bibitem{Ishihara2010}
S. Ishihara, J. Phys. Soc. Japan {\bf 79}, 011010 (2010).

\bibitem{vandenBrink2008}
J. Van den Brink and D. I. Khomskii, J. Phys.: Condens. Matter {\bf 20}, 434217 (2008).

\bibitem{Monceau2012}
P. Monceau, Adv. Phys. {\bf  61}, 325 (2012).

\bibitem{deSouza2013}
M. de Souza and J.-P. Pouget, J. Phys.: Condens. Matter {\bf 25}, 343201 (2013).

\bibitem{STMD2015}
S. Tomi\'c and M. Dressel, Rep. Prog. Phys. {\bf 78}, 096501 (2015).

\bibitem{Abdel-Jawad2010}
M.\ Abdel-Jawad, I.\ Terasaki, T.\ Sasaki, N.\ Yoneyama, N.\ Kobayashi, Y.\ Uesu, and C.\ Hotta, Phys.\ Rev.\ B {\bf 82}, 125119 (2010).

\bibitem{Lunkenheimer2012}
P.\ Lunkenheimer, J.\ M\"uller, S.\ Krohns, F.\ Schrettle, A.\ Loidl, B.\ Hartmann, R.\ Rommel, M.\ de Souza, J.\ A.\ Schlueter  and M.\ Lang, Nat.\ Mater.\ {\bf11} (2012) 755

\bibitem{Tomic2013}
S.\ Tomi\'c, M.\ Pinteri\'c, T.\ Ivek, K.\ Sedlmeier, R.\ Beyer, D.\ Wu, J.\ A.\ Schlueter,D.\ Schweitzer and  M.\ Dressel, J.\ Phys.: Condens.Matter\ {\bf 25} 436004 (2013).

\bibitem{ItohPRL2013}
K.\ Itoh, H.\ Itoh, M.\ Naka, S.\ Saito, I.\ Hosako, N.\ Yoneyama, S.\ Ishihara, T.\ Sasaki, and S. \ Iwai,  Phys.\ Rev.\ Lett.\ {\bf 110}, 106401 (2013).

\bibitem{Pinteric}
M. Pinteri\'c, M. \v{C}ulo, O. Milat, M. Basleti\'c, B. Korin-Hamzi\'c, E. Tafra, A. Hamzi\'c, T. Ivek, T. Peterseim, K. Miyagawa, K. Kanoda, J. A. Schlueter, M. Dressel, and S. Tomi\'c,
Phys.\ Rev.\ B {\bf 90}, 195139 (2014);
M. Pinteri\'c, T. Ivek, M. \v{C}ulo, O. Milat, M .Basleti\'c, B. Korin-Hamzi\'c, E. Tafra,
A. Hamzi\'c, M. Dressel, and S. Tomi\'c,
Physica B {\bf 460}, 202 (2015).

\bibitem{Swietlik2007}
A. Ota, L. Ouahab, S. Golhen, Y. Yoshida, M. Maesato, G. Saito, and R. Swietlik, Chem. Mater. {\bf 19}, 2455 (2007).

\bibitem{Drichko2014}
N. Drichko, R. Beyer, E. Rose, M. Dressel, J. A. Schlueter, S. A. Turunova, E. I. Zhilyaeva, and R. N. Lyubovskaya, Phys.\ Rev.\ B {\bf 89}, 075133 (2014).

\bibitem{ShimizuPRL2003}
Y.\ Shimizu, K.\ Miyagawa, K.\ Kanoda,  M.\ Maesato, and G. \ Saito,  Phys.\ Rev.\ Lett.\ {\bf 91}, 107001 (2003);  Y.\ Kurosaki, Y.\ Shimizu, K.\ Miyagawa, K.\ Kanoda, G.\ Saito, Phys.\ Rev.\ Lett.\ {\bf 95}, 177001 (2005);Y.\ Shimizu, K.\ Miyagawa, K.\ Kanoda, M.\ Maesato, and G.\ Saito, Phys.\ Rev.\ B {\bf 73}, 140407 (2006).

\bibitem{Hotta}
C.\ Hotta, Phys.\ Rev. B {\bf 82}, 241104 (2010); Crystals {\bf 2}, 1155 (2012).

\bibitem{Ishihara}
M.\ Naka and S.\ Ishihara, J.\ Phys.\ Soc.\ Jpn. {\bf 79}, 063707 (2010); J.\ Phys.\ Soc.\ Jpn. {\bf 82}, 023701 (2013).

\bibitem{Sumit}
H.\ Li, R.\ T.\ Clay, and S.\ Mazumdar, J.\ Phys.: Condens.\ Matter {\bf 22}, 272201 (2010);
S.\ Dayal, R.\ T.\ Clay, H.\ Li, and S.\ Mazumdar, Phys.\ Rev.\ B {\bf 83} 245106  (2011).

\bibitem{Gomi}
H.\ Gomi, T.\ Imai, A.\ Takahashi, and M.\ Aihara,
Phys.\ Rev.\ B {\bf 82}, 035101 (2010);
H.\ Gomi, M.\ Ikenaga, Y.\ Hiragi, D.\ Segawa, A.\ Takahashi, T.\ J.\ Inagaki, and M.\ Aihara,
Phys.\ Rev.\ B {\bf 87}, 195126 (2013);
H.\ Gomi, T.\ J.\ Inagaki, and A.\ Takahashi,
Phys.\ Rev.\ B {\bf 93}, 035105 (2016);


\bibitem{Sedlmeier2012}
K.\ Sedlmeier, S.\ Els{\"a}sser, D.\ Neubauer, R.\ Beyer, D.\ Wu, T.\ Ivek, S.\ Tomi\'{c}, J.\ A.\ Schlueter and M.\ Dressel, Phys.\ Rev.\ B {\bf 86}, 245103 (2012).

\bibitem{Kanoda2012}
K. Kanoda, private communication (2012).

\bibitem{DresselRC2016}
M. Dressel, P. Lazi\'c, A. Pustogow, E. Zhukova, B. Gorshunov, J. A. Schlueter, O. Milat, B.Gumhalter,  and S. Tomi\'c,
Phys.\ Rev.\ B {\bf 93}, 081201(R) (2016).

\bibitem{Hiramatsu2016}
T. Hiramatsu, Y. Yoshida, G. Saito, A. Otsuka, H. Yamochi, M. Maesato, Y. Shimizu, H. Ito, Y. Nakamura, H. Kishida, M. Watanabe and R. Kumai, to be submitted.

\bibitem{Shimizu2016}
Y. Shimizu, T. Hiramatsu, M. Maesato, A. Otsuka, H. Yamochi, A. Ono, M. Itoh, M. Yoshida, M. Takigawa, Y. Yoshida, G. Saito, arXiv:1604.01460v1, submitted (2016).

\bibitem{Dressel04}
M. Dressel and N. Drichko, Chem. Rev. {\bf 104}, 5689 (2004); N. Drichko, S. Kaiser, Y. Sun, C. Clauss, M. Dressel, H. Mori,
J. Schlueter, E. I. Zhyliaeva, S. A. Torunova, and R. N.
Lyubovskaya Physica B {\bf 404}, 490 (2009).

\bibitem{Yamamoto05}
T. Yamamoto, M. Uruichi, K. Yamamoto, K. Yakushi, A. Kawamoto, H.
Taniguchi, J. Phys. Chem B {\bf  109}, 15226 (2005).

\bibitem{Girlando11}
A. Girlando, J. Phys. Chem. C {\bf 115}, 19371 (2011).

\bibitem{Kresse93}
G. Kresse and J. Hafner, Phys. Rev. B {\bf 47}, 558 (1993) and
{\bf 48}, 13115 (1993);
G. Kresse and J. Furthm\"{u}ller, Comput. Mat. Sci. {\bf 6}, 15 (1996);
G. Kresse and J. Furthm\"{u}ller, Phys. Rev. B {\bf 54}, 11169 (1996).

\bibitem{Blochl94}
P. E. Bl\"{o}chl, Phys. Rev. B {\bf 50}, 17953 (1994).

\bibitem{Kresse99}
G. Kresse and D. Joubert, Phys. Rev. B {\bf 59}, 1758 (1999)

\bibitem{Bignozzi94}
C. A. Bignozzi, C. Chiorboli, M. T. Indelli, F. Scandola, V. Bertolasi, and G. Cilli,
J. Chem. Soc. Dalton Trans. {\bf 1994}, 2391 (1994).

\bibitem{Togo08}
A. Togo, F. Oba, and I. Tanaka, Phys. Rev. B {\bf 78}, 134106 (2008).

\bibitem{SM}
See Supplemental Material at http://link.aps.org/supplemental/ for more details, and 3D motion picture of the calculated frequencies belonging to the $\nu_{27}$ intramolecular vibrational mode.

\bibitem{Nakamura09}
K. Nakamura, Y. Yoshimoto, T. Kosugi, R. Arita, and M. Imada, J. Phys. Soc. Jpn. {\bf 78}, 083710 (2009).

\bibitem{KandpalPRL2009}
H. C. Kandpal, I. Opahle, Y.-Z. Zhang, H. O. Jeschke, and R. Valent{\'i},
Phys. Rev. Lett. {\bf 103}, 067004 (2009);
H. O. Jeschke, M. de Souza, R. Valent{\'i}, R. S. Manna, M. Lang, and J. A. Schlueter, Phys. Rev. B {\bf 85}, 035125 (2012).

\bibitem{note0}
P$_c$ symmetry implies that while the two (BEDT-TTF)$_2^+$ molecules within each dimer remain symmetrically equivalent, the two dimers are symmetrically inequivalent.

\bibitem{note1}
P2$_1/c$ is the global-scale space group symmetry of the system where all four molecules in the unit cell are equivalent. As opposed to P2$_1$ and P$_c$ it does not allow charge disproportionation to occur.

\bibitem{Saito}
G. Saito et al., unpublished.

\bibitem{Mott1971}
N.\ F.\ Mott and E.\ A.\ Davis, {\em Electronic Processes in Non-crystalline Solids} (Oxford University, London, 1971).

\bibitem{note3}
With a linear shift of 140~cm$^{-1}$ per unit charge, the charge disproportionation $2\delta_\mathrm{\rho}$ is calculated by $2\delta_{\rho} = \delta\nu_{27}/(140~{\rm cm}^{-1}/e)$, where $\delta\nu_{27}$ is the difference in frequency positions between two $\nu_{27}$ vibration modes associated with two non-equal BEDT-TTF sites.

\end{thebibliography}
\end{document}